\documentclass[aps,pre,floatfix,twocolumn,showpacs,amsmath,amssymb,superscriptaddress,titlepage,natbib]{revtex4-1}
\usepackage{graphicx,times}
\usepackage{amssymb,amsmath}
\usepackage{multirow}
\usepackage{color}
\usepackage{epstopdf}
\usepackage{bm}
\usepackage{longtable}
\usepackage{lipsum}

\begin{document}

\title{The effects of noise and time delay on the synchronization of the Kuramoto model in small-world networks}

\author{Sara Ameli}
\affiliation{ Department of Physics, Isfahan University of Technology, Isfahan 84156-83111, Iran}

\author{ Farhad Shahbazi} \email{shahbazi@cc.iut.ac.ir }
\affiliation{ Department of Physics, Isfahan University of Technology, Isfahan 84156-83111, Iran}

\author{ Maryam Karimian}
\affiliation{ Department of Physics, Isfahan University of Technology, Isfahan 84156-83111, Iran}
\affiliation{ Maastricht Centre for Systems Biology, Maastricht University, Maastricht, 6229ER, The Netherlands}

\author{ Tahereh Malakoutikhah}
\affiliation{ Department of Physics, Isfahan University of Technology, Isfahan 84156-83111, Iran}


\begin{abstract}
We study the synchronization of a small-world network of identical coupled phase oscillators with Kuramoto interaction. 
First, we consider the model with instantaneous mutual interaction and the normalized coupling constant to the degree of each node. 
For this model,  similar to the constant coupling studied before, we find the existence of various attractors corresponding to the different defect patterns and also the noise enhanced synchronization when driven by an external uncorrelated white noise.  We also investigate the synchronization of the model with homogenous time-delay in the phase couplings. For a given intrinsic frequency and coupling constant, upon varying the time delay we observe the existence a partially synchronized state with defect patterns which transforms to 
 an incoherent phase characterized by randomly phase locked states. By further increasing of the time delay, this phase again undergoes a  transition to another patterned partially synchronized state. We show that the transition between theses phases are discontinuous and moreover in each phase the average frequency of the oscillators decreases by increasing the time delay and shows an upward jump at the transition points.  
\end{abstract}

\maketitle

\section{Introduction}

Synchronization is a manifestation of the coherence in dynamical behavior in a population of interacting units with sustainable oscillations. Phase synchrony in the dynamics of coupling self-sustained oscillators has been studied widely for modeling of many natural collective phenomena including chemical, biological and neuronal systems~\cite{winfree2001geometry,pikovsky2003synchronization, manrubia2004emergence, balanov2008synchronization}. The Kuramoto model, introduced by Y. Kuramoto~\cite{kuramoto1975self,kuramoto2012chemical},  is a minimal model describing a set of phase oscillators, in which each pair of oscillators is mutually coupled through an odd  2$\pi$ periodic function of their phase differences. 
This model is exactly solvable for an all-to-all network and shows a continuous  phase transition from incoherent dynamical state to a synchronized state at a critical coupling strength~\cite{acebron2005kuramoto}. 

The recent rapid growth in network sciences motivated much research on the dynamical systems in complex networks~\cite{boccaletti2006complex}. Consequently,  synchronization ~\cite{arenas2008synchronization} and particularly the Kuramoto model  in complex networks has been studied extensively~\cite{rodrigues2016kuramoto}.  
In the real world, the network topology of interacting units is neither completely regular nor random. The small-world (SW) network, proposed by Watts and Strogatz~\cite{watts1998collective, watts2001small} has bilateral features, the high degree of clustering as in regular networks and the small average distance like the random networks. There have been a few studies on the synchronization of the Kuramoto model in  SW networks~\cite{gade2000synchronous,watts2001small,hong2002synchronization, barahona2002synchronization,  wang2002synchronization,esfahani2012noise}. In an attempt to discover the effect of SW topology on the synchronization of Kuramoto model, it is found that for an identical group of
phase oscillators coupled by a constant coupling strength,  there are a variety of attractors~\cite{esfahani2012noise}. In other words, starting from different random initial phase distributions, the stationary state of the systems would be different. It was shown that the system in the stationary states is partially synchronized due to the presence of some defect patterns. These are either the point defects, i.e   the isolated phase textures continuously varying from $0$ to $\pi$, or the quasi-periodic patterns in which the phase of the oscillators varies almost periodically in the systems. 
These defects are shown to be stable against small local perturbation, hence are topologically protected~\cite{esfahani2012noise}. Moreover, it was found that introducing an additive white noise to the Kuramoto dynamics as an external force, causes the elimination of the defects at some specific noise strength, leading to the growth of synchrony among the oscillators.  This phenomenon is called  {\em stochastic synchronization} and is a peculiarity of SW networks and was not observed in other well known complex networks such as random and scale-free networks~\cite{khoshbakht2008phase}.

Another interesting issue in the problem of synchronization is the effect of time delay in the interaction between the oscillators. 
Most often,  the interaction between the oscillating units in a network is not instantaneous, and there is a delay due to a finite signal transmission that can not be neglected. One example is the nervous system, in which the delays are $10$-$200$ ms, the same order as the signal duration~\cite{buzsaki2004neuronal, nunez2006electric}. The limit cycle oscillators coupled with time delayed Kuramoto interaction have been studied for two mutually coupled oscillators~\cite{schuster1989mutual}, a square lattice of nearest neighbor coupled oscillators~\cite{niebur1991collective} and also all-to-all networks~\cite{yeung1999time,choi2000synchronization} with homogenous time delay.
For both identical and non-identical oscillators with a given coupling constant and time delay, non-unique stable solutions with different collective frequencies are possible~\cite{schuster1989mutual,niebur1991collective,choi2000synchronization}.  The phase diagram of the time delayed Kuramoto model for a fully connected network of both identical and non-identical  oscillators with randomly distributed natural frequencies, reveals three phases. These are coherent, incoherent and a bi-stable phase in which both the coherent and incoherent solutions coexist with the possibility of time-periodic solutions for the order parameter~\cite{yeung1999time}. The dispersion in synchronization frequency of a network of sparsely connected time delayed oscillators with non-identical degree distributions has also been observed~\cite{nordenfelt2014frequency}.

We organize the paper as the following. In section~\ref{noise}, we investigate the effect of normalization of coupling constant by the degree of each node on the dynamics of the Kuramoto model in an SW network of identical phase oscillators. We show that the normalization of the coupling constant does not have a remarkable effect on the defect patterns and also the presence of stochastic synchronization, except the lowering of the ratio of noise intensity to the coupling constant at which the noise enhancement of the synchronization occurs. 
In section~\ref{delay}, we show that the defects are also stable against small time delay. Even Interestingly, the intermediate values of time delays could increase the synchronization by weakening the defects. We also discuss the possible phases and transitions between them as the time delay varies.  Section \ref{conclusion} is devoted to the concluding remarks.

\section{The Kuramoto model with normalized couplings in a small-world network  }
\label{noise}

\subsection{Deterministic case}

Consider a set of $N$ phase oscillators with intrinsic frequencies, $\omega_i$,  $i=1, ..., N$. In terms of  network terminology, each oscillator is considered  as a node and the mutual communication between  any two oscillators is demonstrated by an edge between them.   In Kuramoto model  the dynamics of this system is given by the following set of coupled equations~\cite{kuramoto1975self}:
\begin{equation}
\label{kuramoto1}
\frac{d{\theta}_i}{dt}=\omega_i+\frac{K}{k_{i}}\sum^{N}_{j=1} a_{ij}\sin(\theta_j-\theta_i),
\end{equation}
where $\omega_{i}$, $\theta_i$ and $k_i$ denote the intrinsic frequency, phase and the degree (number of neighboring nodes) of the oscillator at  node $i$, respectively.  $\frac{K}{k_i}$ is the normalized coupling constant  and $a_{ij}$ denotes the matrix element of adjacency matrix($a_{ij}=1$ if nodes $i$ and $j$ are linked together and $a_{ij}=0$ otherwise).

In this work, we consider small-world topology for the network consisting of similar  oscillators with $\omega_{i}=\omega_{0}$. Therefore, it is possible to eliminate $\omega_{0}$ from equation \eqref{kuramoto1} by choosing  a rotating framework, i. e $\theta'_{i}=\theta_{i}-\omega_{0}t$, in which this equation can be rewritten as  

\begin{equation}
\label{kuramoto2}
\frac{d{\theta}_i'}{dt'}=\frac{1}{k_{i}}\sum^{N}_{j=1} a_{ij}\sin(\theta'_j-\theta'_i),
\end{equation}
where $t'=Kt$ is the rescaled time. 

To construct a small-world network we use Watts-Strogatz (WS) algorithm~\cite{watts1998collective,watts2001small} which gives the two main small-world features, small mean path length and large clustering coefficient, to a regular lattice by random rewiring of  its edges provided that the rewiring  probability be $0.005 \lesssim p \lesssim 0.05$.   
In this work we perform our calculations on  a WS network of $N=1000$ nodes and average degree $<k>=10$, by choosing  the  rewiring probability $p=0.03$.

To numerically investigate the synchronization  of Kuramoto model, we integrate equation~\eqref{kuramoto1}  to obtain the time evolution of $\theta_i$, starting from a uniformly random  initial phase distribution  in the interval $[-\pi,\pi]$. For the numerical integration we use the forth order Runge-Kutta method with the time step $dt=0.01$.

To quantify the degree of synchronization, we define the following order parameter:
\begin{equation}
r=\frac{1}{N}|\sum^N_{j=1}e^{i\theta_j}|
\end{equation}
The order parameter $r$  vanishes in the cases of  randomly distributed (unsynchronized) or phase locking with regular phase lag, and equals  to $1$ when the phases are  all the same (full synchrony). The partial synchronized states are characterized by $0 < r < 1$. The order parameter provides information about global phase distribution, henceforth to  gain insight into the local phase configurations of the system we also calculate  the correlation matrix $D$ defined as
\begin{equation}
D_{ij}= \lim_{\Delta t \to \infty}\frac{1}{\Delta t}\int^{\Delta t+t_r}_{t_r}\cos(\theta_i(t)-\theta_j(t))dt.
\end{equation}
where $t_r$ is the time needed for reaching to stationary state~\cite{gomez2007paths}. Here we used $5\times 10^5$ time steps for the stationary time $t_r$ and $3\times 10^4$  time steps for the averaging window $\Delta t$.  The correlation matrix element $D_{ij}$ is a measure of coherency between the pair of oscillators at $i$ and $j$ positions and takes a value in the  interval $1\leq D \leq -1$. $D_{ij}=1$ when there is a full synchrony between nodes $i$ and $j$ ($\theta_i=\theta_j$), and $D_{ij}=-1$ when they are in the anti-phase state ($\theta_i=\theta_j \pm \pi$).

\begin{figure}[t]
\centering
\includegraphics[width=\columnwidth]{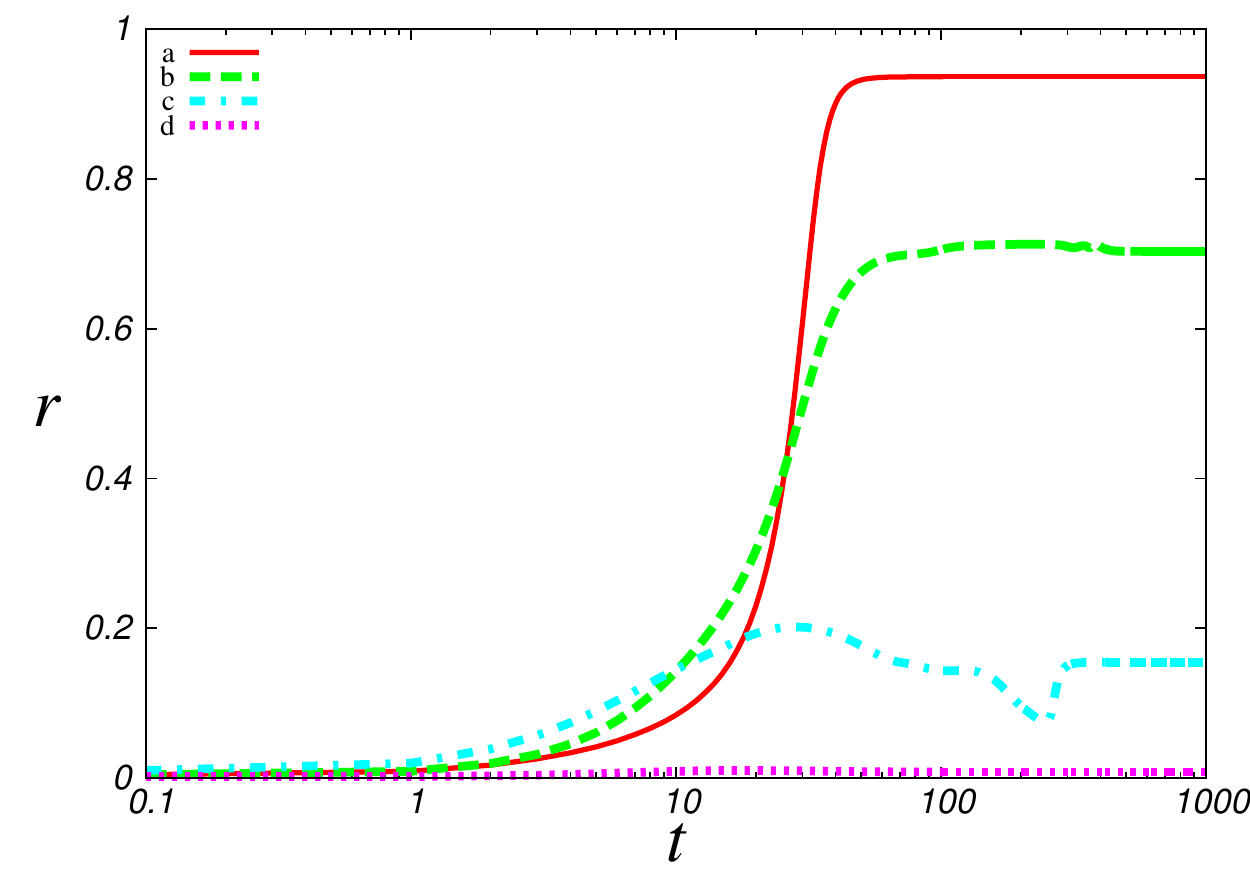}
\caption{(Color online) Order parameter (r) versus time (in logarithmic scale) for WS networks  with $N = 1000$ and $<k> = 10$, for four different initial conditions.}
\label{r-t}
\end{figure}

\begin{figure}[h]
\includegraphics[width=\columnwidth]{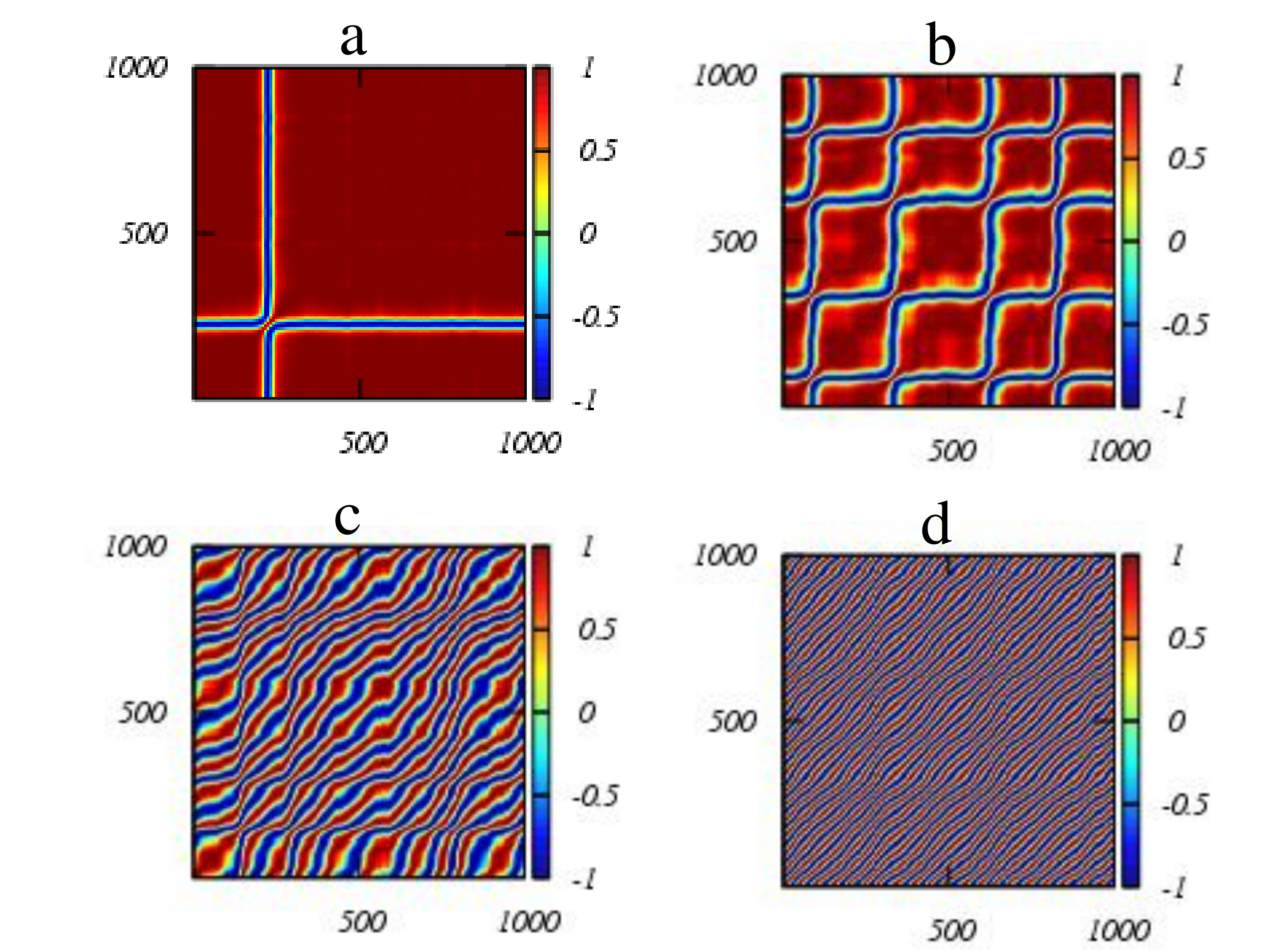}
\caption{(Color online) Density plot of correlation matrix elements ($D_{ij}$ )  corresponding to the four stationary states shown in 
figure~\ref{r-t}.}
\label{D-sw}
\end{figure}

The time dependence of the order parameter $r$, is represented in figure~\ref{r-t}, showing that the stationary state of system at long time limit depends on initial phase distribution. As shown in this figure the four different initial condition leads to four different steady states. This result has also been obtained for the Kuramoto model without coupling normalization~\cite{esfahani2012noise}. 
The density plots of the correlation matrix, corresponding to initial conditions in Fig.~\ref{r-t}, are represented in Fig.~\ref{D-sw}. The interesting result deducing from this figure, is that the defect patterns created in the steady  phase configuration are exactly the same as what were found before for the case of  non-normalized  constant~\cite{esfahani2012noise}. This results show that such defects  are resistant against the normalization of the coupling constants by the node degrees.

\subsection{Noisy Kuramoto model}

 A network of oscillators might be affected  by some external random forces which also can be assigned to the temporal fluctuations of intrinsic frequencies. To take such a random fluctuations into account, we introduce  an uncorrelated white noise to the Kuramoto model.  A white noise has  zero mean value $\langle\eta(t) \rangle=0$ and the spatiotemporal correlation given by:
 
\begin{equation}
\langle\eta_\alpha(t_1)\eta_\beta(t_2) \rangle=2D\delta(t_1-t_2)\delta_{\alpha \beta},
\end{equation}
where $D$ is the variance or intensity of the noise. Introducing  such a  noise to the model, one finds

\begin{equation}\label{kn}
\frac{d\theta_i'}{dt}=\frac{K}{k_i}\sum^N_{j=1}a_{ij}\sin(\theta_j'-\theta_i')+\eta_i(t).
\end{equation}
Rescaling the time as $t'=K t$ and the noise as $\eta_{i}(t')=\sqrt{KD}\xi_{i}(t')$, we arrive at

\begin{equation}\label{noisy-k}
\frac{d\theta_i'}{dt'}=\frac{1}{k_i}\sum^N_{j=1}a_{ij}\sin(\theta_j'-\theta_i')+g\xi_i(t'),
\end{equation} 
where $g=\sqrt{\frac{D}{K}}$ is a dimensionless parameter measuring the ratio of  noise intensity  to the coupling constant and $\xi(t')$ is  spatially uncorrelated  Gaussian white noise with the unit variance  

\begin{equation}
\langle\xi_\alpha(t_1')\xi_\beta(t_2') \rangle=2\delta(t_1'-t_2')\delta_{\alpha \beta}.
\end{equation}

Now, using Ito stochastic integration~\cite{gardiner1985handbook}, the equation \eqref{noisy-k} can be integrated to find out its long-time features. The quantity of interest, characterizing  the synchronization in the stationary state of the system is the mean value of the order parameter at long time limit, i.e 
\begin{equation}
r_{\infty}=\lim_{t_{r}\rightarrow \infty} \lim_{\Delta t \rightarrow\infty} \frac{1}{\Delta t}\int_{t_r}^{t_r+\Delta t} r(t) dt . 
\end{equation}

\begin{figure}[t]
\centering
\includegraphics[width=0.7\columnwidth, angle=270]{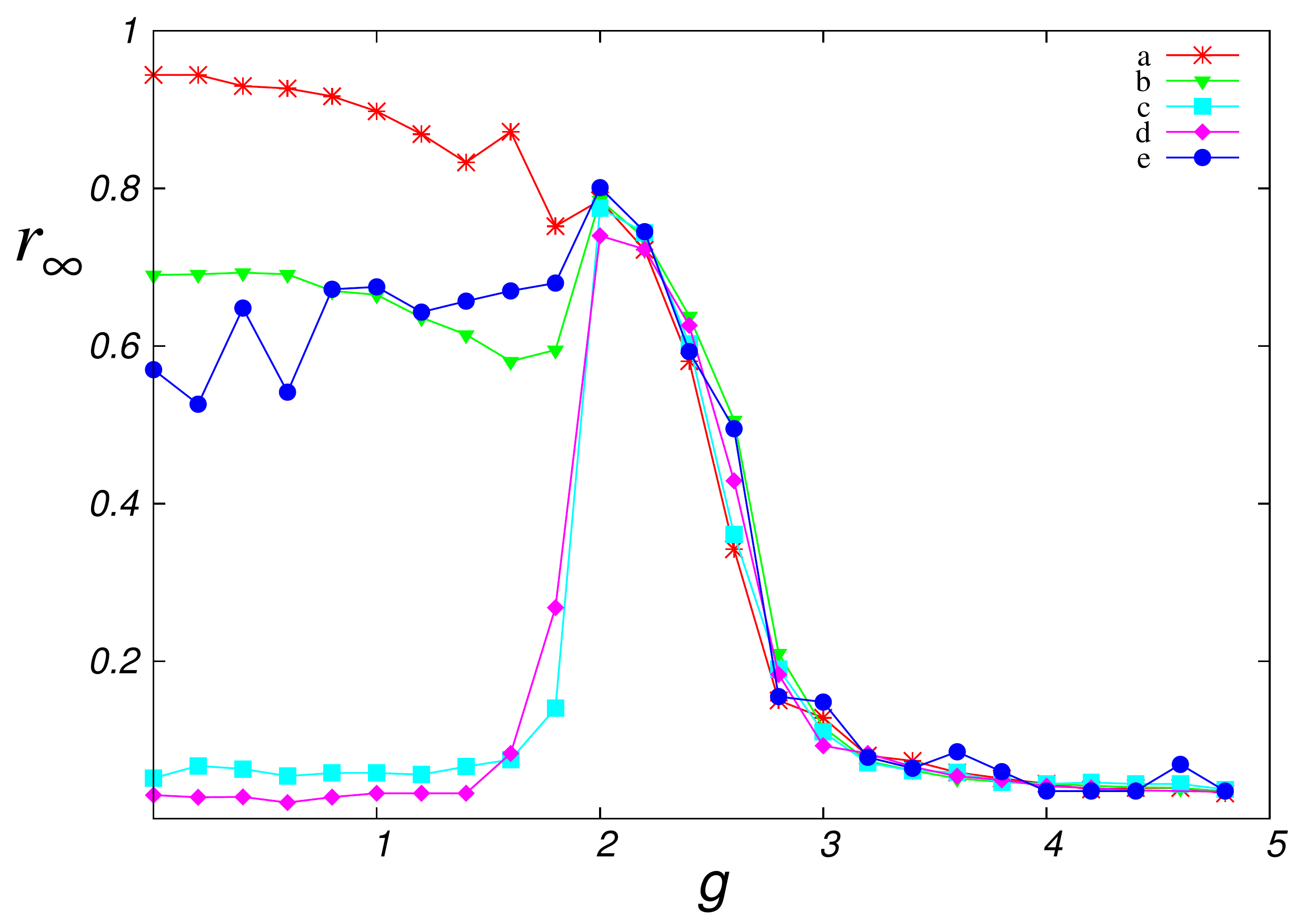}
\caption{(Color online) The stationary order parameter ($r_{\infty}$) versus $g$ for a WS network of $N=1000$ oscillators corresponding to the  five different initial conditions.}
\label{r-g}
\end{figure}
\begin{figure}
\includegraphics[angle=270,width=0.8\columnwidth]{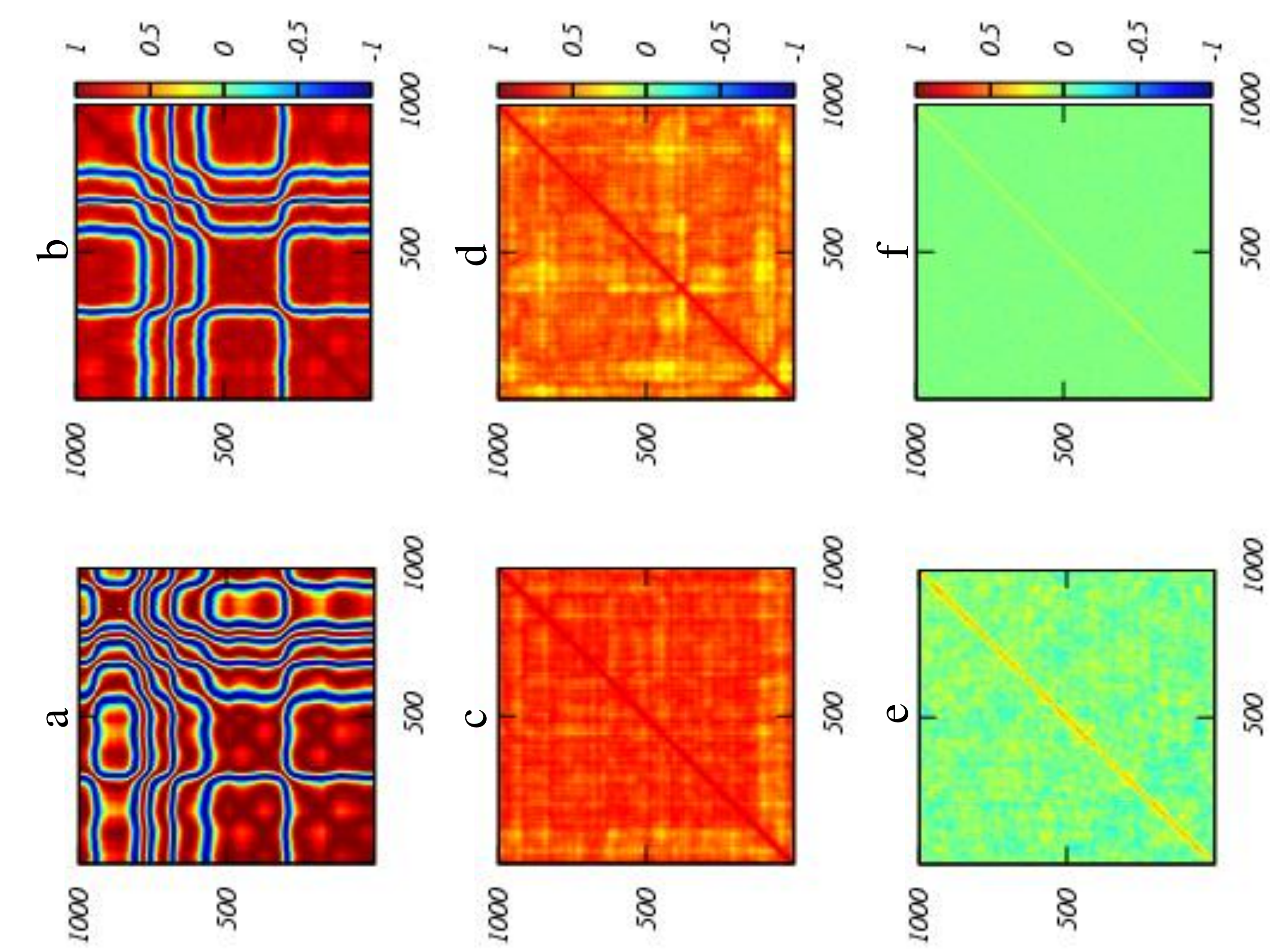}
\centering
\caption{(Color online) Density plot of the elements of stationary correlation matrix  for a WS network of $N=1000$ with random initial phase distribution in the presence of a driving noise with  (a) $g=0$, (b) $g=1$, (c) $g=2$, (d) $g=2.2$ , (e) $g=3$, (f) $g=4$.}
\label{D-all}
\end{figure}

Fig.~ \ref{r-g} represents the dependence of $r_{\infty}$ with respect to reduced parameter $g$, in a WS network consisting of $N=1000$ nodes and in the case that noise is applied on all of the oscillators. Different plots in this figure correspond to the different initial phase distributions, giving rise to distinct defect  patterns. Fig. \ref{r-g} shows that starting from the most of initial conditions the synchrony among the oscillators increases  as the ratio of noise intensity to the coupling constant reaches the values in such a way that $g$ lays in a narrow range around $g\sim 2$. The onset of such a noise enhanced synchronization is lower  for the Kuramoto model with normalized coupling with respect to the constant coupling case~\cite{esfahani2012noise}, where  the enhancement in synchronization occurs at $g\sim 6$.  As pointed out by Kouhi et al.~\cite{esfahani2012noise} , the reason for such a  counter-intuitive increase in the synchrony  is the diminishing of the defects as the effect of noise at a critical value of the noise strength. This can clearly be seen in the density plots of the correlation matrix in figure \ref{D-all}, which illustrate that the defect patterns totally vanishes at $g\sim2$  and so the systems settles down in a  homogeneous state.   Noise intensities greater than  $g\sim 2$, where there is no more defects, have  destructive effect on the synchronization.

\section{effect of the time delay}
\label{delay}

\begin{figure}[b]
\centering
\includegraphics[width=0.9\columnwidth]{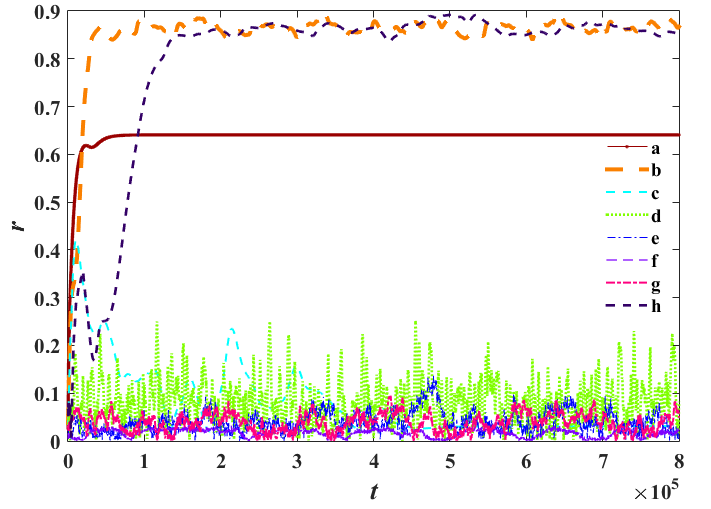}
\caption{(Color online) The time dependence of the order parameter for WS network with $N=1000$ and $<k>=10$ with a   random initial phase distribution, for  the phase shifts $\omega_0\tau$ equal to (a) 0, (b) 0.94, (c) 1.41, (d) 1.73, (e) 1.88, (f) 2.35, (g) 4.082, (h) 4.24.}
\label{r-t-tau}
\end{figure}
\begin{figure*}
\includegraphics[trim = 10mm 0mm 0mm 0mm, clip, width=1.05\textwidth]{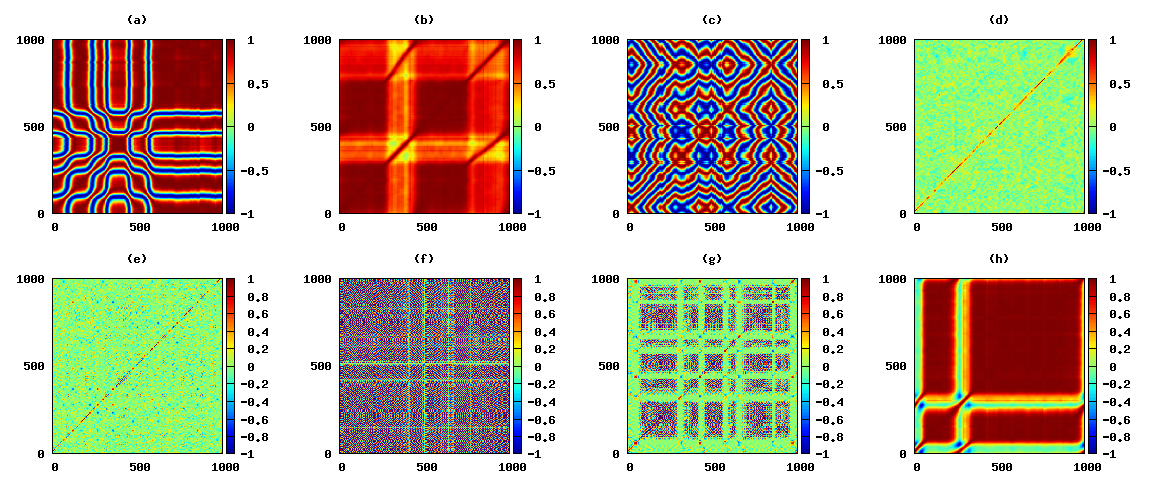}
\caption{(Color online) Density plot of the correlation matrix elements  the phase shifts $\omega_0\tau$ equal to (a) 0, (b) 0.94, (c) 1.41, (d) 1.73, (e) 1.88, (f) 2.35, (g) 4.082, (h) 4.24.}
\label{Forward1}
\end{figure*}
\begin{figure*}
\includegraphics[trim = 0mm 0mm 0mm 0mm, clip, width=1\textwidth]{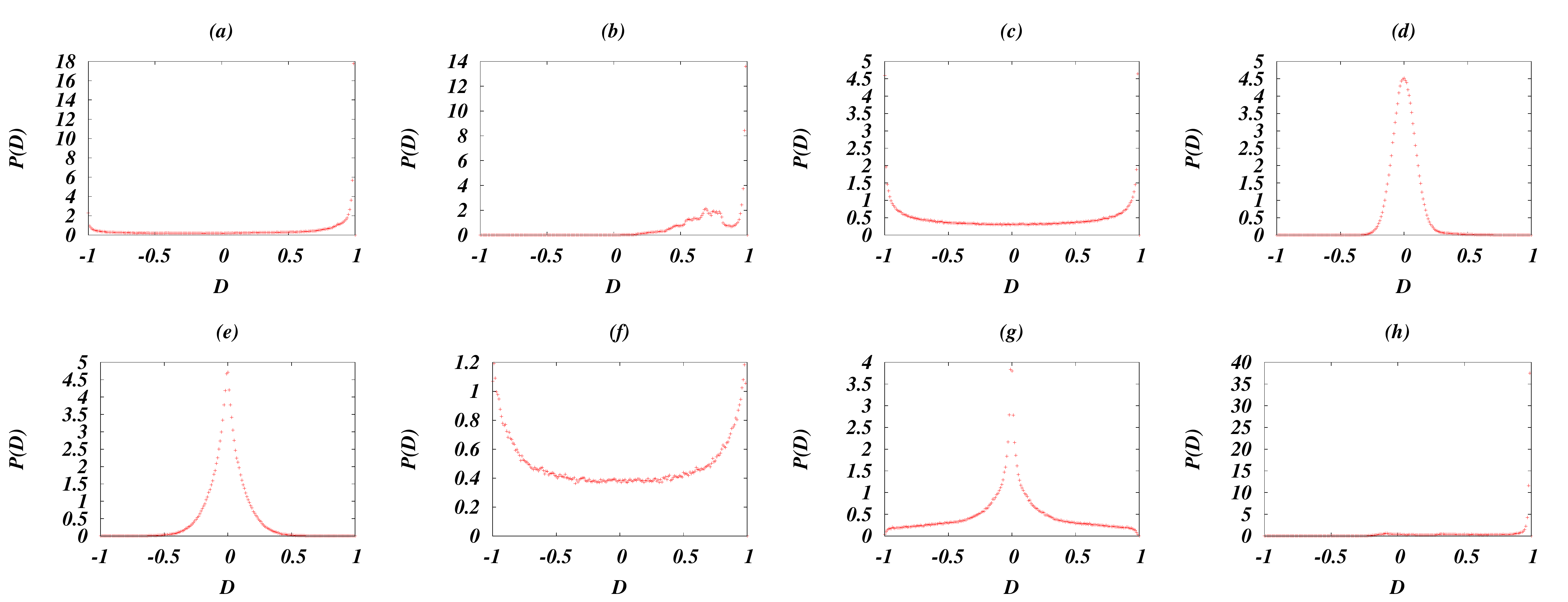}
\caption{(Color online)  The probability density function of correlation matrix elements for the phase shifts $\omega_0\tau$ equal to (a) 0, (b) 0.94, (c) 1.41, (d) 1.73, (e) 1.88, (f) 2.35, (g) 4.082, (h) 4.24.}
\label{Forward2}
\end{figure*}
\begin{figure*}
\centering
\includegraphics[width=\textwidth]{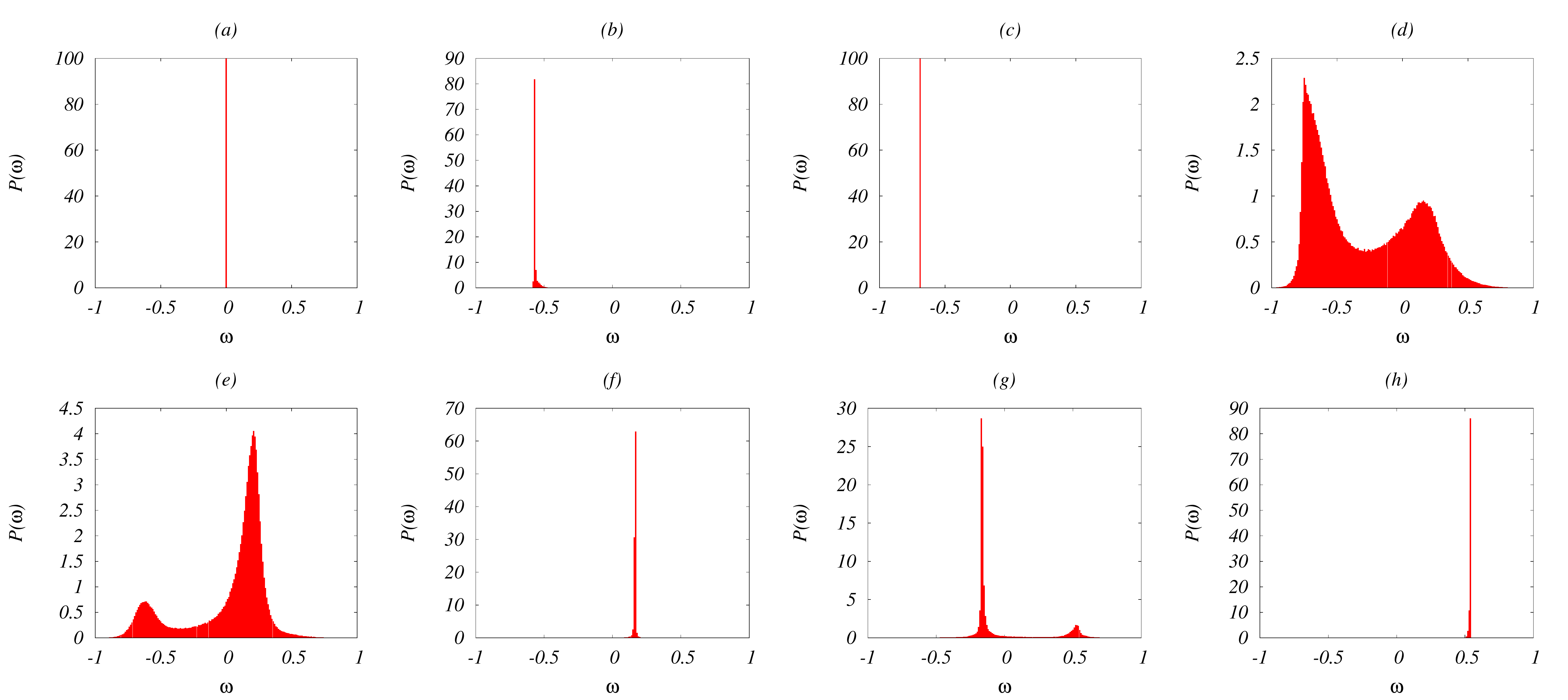}
\caption{(Color online) The probability density function of frequency of the oscillators for the phase shifts $\omega_0\tau$ equal to (a) 0, (b) 0.94, (c) 1.41, (d) 1.73, (e) 1.88, (f) 2.35, (g) 4.082, (h) 4.24.}
\label{pw}
\end{figure*}
\begin{figure}
\centering
\includegraphics[width=\columnwidth]{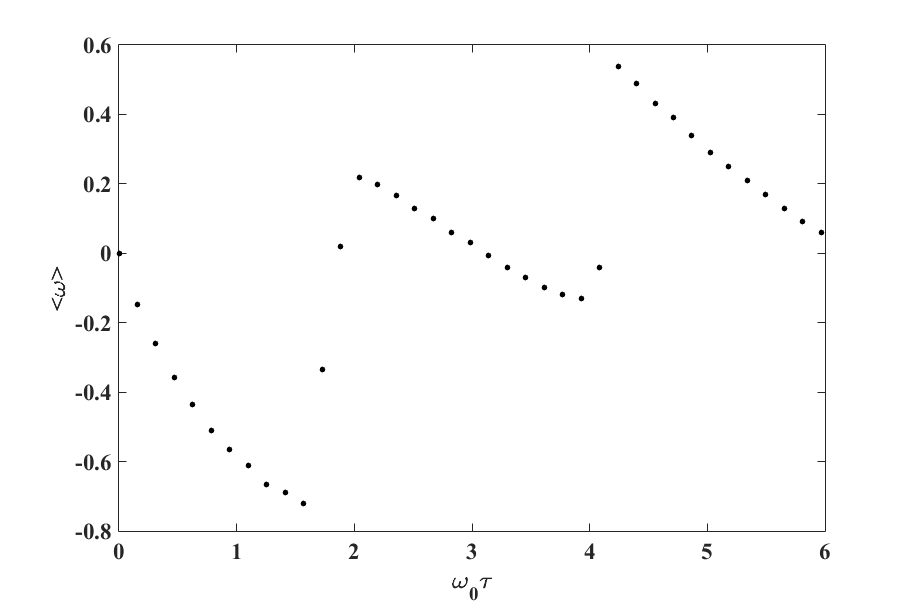}
\caption{Long-time averaged angular frequency of oscillators, $\langle\omega\rangle$,  versus the phase shift $\omega_{0}\tau$. }
\label{wtau}
\end{figure}

\begin{figure}
\centering
\includegraphics[width=\columnwidth]{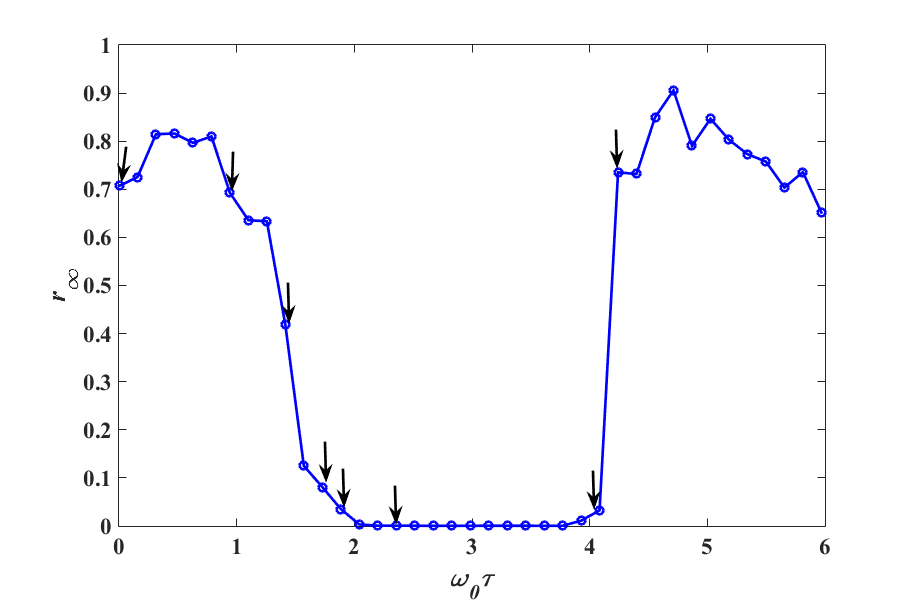}
\caption{(Color online) Long-time averaged order parameter ($r_{\infty}$)  versus the time delay. The arrows indicate the phase shift  corresponding to those shown in figures \ref{Forward1}, \ref{Forward2} and \ref{pw}.}
\label{Rtau}
\end{figure}

In this section we investigate the effect the time delay on the synchronization of deterministic Kuramoto model.  For simplicity we assume a uniform time delay for all the coupled oscillators, so we have  

\begin{equation}\label{delay1}
\frac{d\theta_i}{dt}=\omega_0+\frac{K}{k_i}\sum^N_{j=1}a_{ij}\sin \left(\theta_j(t-\tau)-\theta_i(t) \right),
\end{equation}
where $\tau$ is the time delay between  any connected pair of oscillators. 
Defining ${\theta}'_i(t')=\theta_i(t)-\frac{\omega_0}{K}t' $ with $t'=Kt$, equation \eqref{delay1} is turned  to the following form  

\begin{equation}\label{delay2}
\frac{d\theta'_i(t')}{dt'}=\frac{1}{k_i}\sum^N_{j=1}a_{ij}\sin\left(\theta'_j(t'-\tau')-\theta'_i(t')-\frac{\omega_0}{K}\tau'\right).
\end{equation}
Therefore, unlike the instantaneous coupling, in the presence of the time delay it is not possible to eliminate the intrinsic frequency $\omega_{0}$ by moving to a rotating frame. In this case the results depend on the intrinsic frequency and the coupling constant through their ratio $\omega_{0}/K$. Hence, from now on, we choose $K=1$ and  represent the results in terms of  $\omega_{0}\tau$ which appears as a phase shift in equation~\eqref{delay2}. 

In figure \ref{r-t-tau}, we represent the  time dependence of the order  for some values of the phase shift, i.e. $\omega_{0}\tau=0, 0.94, 1.41, 1.73, 1.88, 2.35, 4.082, 4.24$. This figure shows that upon increasing the time delay, the coherency in the system first increases and then falls for a long period and then again rises. To gain insight into the local dynamics, for each of the phase shifts mentioned above,  we calculate the long time averaged correlation matrix as well as the probability density function of its elements, shown in figures \ref{Forward1} and \ref{Forward2}, respectively. 

Figure \ref{Forward1} unveils  three distinct  collective behavior: (i) the partially synchronized states with visible point-like
 (Figs.\ref{Forward1}-(a), (b), (h)) or quasi-periodic pattern   (Fig.\ref{Forward1}-(c));
(ii) the dynamic incoherent  states (Fig.\ref{Forward1}-(d)), where the fluctuating phase differences between any pair of oscillators vanishes their mutual phase correlation, hence there is no phase locking in this state; 
and (iii) the static incoherent or random phase locked states (Fig.\ref{Forward1}-(f)), in which any two oscillators rotate with an almost constant phase difference but the amount of this phase difference is completely random for  different pairs of oscillators.  In this state, there is a random freezing in phase differences, and so there is no distinguishable pattern, which resembles the spin freezing in the spin glass materials. Moreover, the coexistence of static and dynamic incoherent states is possible (Figs.\ref{Forward1}-(e), (g)). 

In terms of the probability density of the correlation matrix elements (Fig.~\ref{Forward2}), the partially localized states with isolated defects are characterized by bimodal PDF's reflecting a majority of synchronized  oscillators coexisting with minor anti-phase ones known as defects (Figs. \ref{Forward2}-(a) , (b), (h)). 
However, for the quasi-periodic patterns, a uniform distribution of the phase differences is observed (Fig.\ref{Forward2}-(c)).  Note that the uniform distribution for the phase difference $\Delta\theta$ between $0$ and $\pi$ gives rise to diverging peaks at $-1$ and $+1$ and flat curve in between for the probability density function of  $\cos(\Delta\theta)$. 
The probability density function  for the dynamic incoherent states is a Gaussian centered at zero, meaning the absence of correlation in the phase dynamics (Fig.\ref{Forward2}-(d)), and an approximate uniform phase difference distribution for  the random phase locked  state  (Fig.\ref{Forward2}-(f)), and 
sharply peaked at zero with heavy tails for the coexistence of the two incoherent states (Figs.\ref{Forward2}-(e) (g)). 

To gain more insight into the collective dynamics of the three phases, we calculate the probability density functions for the frequency of the oscillators in the rotating frame, that is $\omega_{i}=\dot\theta_{i}-\omega_{0}$, after reaching the system to the steady state and over  a long time period  ($10^5$ time steps). 
The results for the values of phase shift mentioned before are displayed in figure~\ref{pw}. It can be seen that in the absence of time delay, $p(\omega)$ has a sharp peak at $\omega=0$ (Fig.\ref{pw}-(a)), which means that in the steady state all the oscillators rotate with the intrinsic frequency $\omega_0$. By increasing the time delay this peak first  
shifts leftward and show small dispersion (Fig.\ref{pw}-(b)) unless for the quasi-periodic patterns where there is no dispersion (Fig.\ref{pw}-(c)).  Interestingly, for the dynamic incoherent states (Fig.\ref{pw}-(d), (e)) $p(\omega)$ is a bimodal distribution with two peaks at negative and positive angular frequencies. This indicates that in the dynamic incoherent states the oscillators, rotate with two dominant frequencies, one less and the other more than $\omega_0$,  preventing the system to reach a synchronized state. As a result of such large dispersion in the frequency distribution, the order parameter represents rapid oscillations (see Fig.\ref{r-t-tau}).  Further increasing of the time delay causes a transition to the random phase-locked state, in which $p(w)$ illustrate a sharp peak with small width at a positive frequency (Fig.\ref{pw}-(f)). At the second transition point, another peak appears at a higher frequency (Fig.\ref{pw}-(g)) and finally the systems sets in a larger positive frequency at the next partially coherent phase (Fig.\ref{pw}-(h)). The dependence of the collective frequencies on the phase shift is illustrated in figure \ref{wtau}, in which $\langle \omega\rangle=\int \omega p(\omega) d\omega$ is the average frequency. This figure clearly shows the decrease of the average frequency by the time delay at each phase and also rapid jump at the transition between the partially coherent and incoherent phases. 

To wrap up this section we discuss the long-time averaged order parameter ($r_{\infty}$) depicted in figure \ref{Rtau}, which represents the stationary order parameter averaged over $14$ different initial phase distributions.  This figure shows that in the partially synchronized, where $0 < \omega_{0}\tau \lesssim \pi/2$, $r_{\infty}$ does not vary  monotonically versus the time delay. Indeed, by increasing the time lag $\tau$, the synchronization in the system in average grows initially and after reaching to a maximum, begins to decreases
and ends up with a rapid fall at $\omega_{0}\tau\sim \pi/2$ to the random phase locking state. It remains exactly zero in a randomly locked state for $\pi/2 \lesssim \omega_{0}\tau \lesssim 3\pi/2$, and then again at $\omega_{0}\tau \sim 3\pi/2$ shows a sudden upward jump to a partially synchronized state with higher collective frequency,  where one  observes a first upward and then a downward behavior  in the order parameter up to $\omega_{0}\tau \sim 2\pi$. 

We can explain the observed collective dynamics of the systems as the following. For $K=1$, the time derivative of each phase in the rotating frame given by equation \ref{delay2} can be rewritten as

\begin{eqnarray}\label{delay3}
\frac{d\theta'_i(t)}{dt}=&\frac{1}{k_i}\sum^N_{j=1}a_{ij}\sin[\theta'_j(t-\tau)-\theta'_i(t)]\cos{\omega_0}\tau \nonumber\\
-&\frac{1}{k_i}\sum^N_{j=1}a_{ij}\cos[\theta'_j(t-\tau)-\theta'_i(t)]\sin{\omega_0}\tau. \nonumber\\
\end{eqnarray} 
For $\omega_{0}\tau< \pi/2$ both the coefficient of $\sin$ and $\cos$ terms in right hand side of equation~\ref{delay3}  are positive. With positive coupling, the $\sin$ term stabilizes  the coherence between any pair of mutually interacting  oscillators, then the first term tends to decrease the phase difference $\delta\theta'=\theta'_j(t-\tau)-\theta'_i(t)$. As a result, the $\cos(\delta\theta')$ in second term would be positive, which causes the decreasing of   the frequency $\frac{d\theta'_i(t)}{dt}$ as  the phase shift increases up to $\omega_{0}\tau=\pi/2$. However, for $\pi/2 < \omega_{0}\tau < 3\pi/2$, the coefficient of $\sin(\delta\theta')$  is negative, therefore this term encourages a phase difference equal to $\pi$. Because of the large clustering in the small-world networks, this gives rise to a frustration and together with the disorder originating from non-homogeneity in  connectivity of the nodes, tends to random freezing of the phase differences. The jump in the frequency at $\omega_{0}\tau\sim\pi/2$, is then due to the sign change of  $\cos(\delta\theta')$ as the $\delta\theta'$ switches from $0$ to $\pi$ at this point.  At $\omega_{0}\tau=3\pi/2$, the sign of the coefficient of $\sin(\delta\theta')$ becomes positive and hence this term again tends to stabilize the coherency in the system,  which is accompanied with an upward jump in the frequency. This is  because $\cos\omega_{0}\tau < 0$  and  as the phase difference shifts from $\pi$ to $0$, the sign of the second term changes from negative to positive.

\section{conclusion}
\label{conclusion}

In summary, we studied the Kuramoto dynamic for a small-world network a similar phase oscillators with normalized coupling constants. 
We found that like the model with uniform coupling constant, the defect structures appear in this model and remain resistant in against the presence of noise. The stability of these patterns against both deformation of the  couplings reveals their topologic nature. It is also shown that the noise of specific intensity can destroy the defects and hence gives rise to noise enhanced synchronization effect in this model. 

We also investigated the effect of the time delay in this model and observed that variety of complex behaviors appears in the collective dynamics of the model  in the  case of homogenous time delay.
Such a time delay can drive the system toward a more homogenous state by destroying  the defects and hence enhances the synchronization. The enhancement of synchronization as the effect of time delay has also been reported in a group of neuronal oscillators~\cite{PhysRevLett.92.074104, PhysRevLett.92.114102}. The coherent states coexist with the phase uncorrelated  incoherent states and undergo a discontinuous  transition to a randomly quenched phase state. Moreover, this transition is accompanied by a discontinuous jump in the average frequency of  oscillators, which is similar to what is found in  all-to-all network~\cite{choi2000synchronization}. 
The discontinuities in both the collective frequency and order parameter at transition from partially synchronized to incoherent quenched phase locked state are the indications  for the  realization of {\em explosive synchronization}  in time delayed Kuramoto model in small-world networks.  The explosive synchronization which is the discontinuous transition in the synchrony of a group of oscillators  is seen in the scale-free networks with the intrinsic frequencies  proportional to the degree of the nodes~\cite{gomez2011explosive}. The enhancement of the explosive  synchronization has also been observed as the  effect of the time delay~\cite{PhysRevE.86.016102}.

This work is an evidence for the extreme complexity of small-world topologies in which even an identical group of interacting phase oscillators represents novel collective phenomena. The full phase diagram of the model with time delay coupling requires more investigation and is left for future studies. We also hope to investigate  the possible applications of  this work in the synchronization of neurons in small-world brain networks~\cite{bassett2006small}.

\begin{acknowledgements} 

The authors gratefully acknowledge the Sheikh Bahaei National High Performance Computing Center (SBNHPCC) for providing computing facilities and time. SBNHPCC is supported by scientific and technological department of presidential office and Isfahan University of Technology (IUT).
\end{acknowledgements}

\bibliographystyle{apsrev4-1}

\bibliography{bibliography.txt}

\end{document}